\documentclass[a4paper,aps,pra,preprintnumbers]{revtex4}
\usepackage[utf8]{inputenc}
\usepackage[english]{babel}
\usepackage[T1]{fontenc}
\usepackage{amssymb,amsfonts,amsmath,mathtext,enumerate,float,dsfont}
\usepackage{graphics,graphicx,epsfig,epstopdf}
\usepackage{caption}
\usepackage{cmap}
\usepackage{multirow}
\usepackage{indentfirst}
\usepackage[usenames]{color}
\usepackage{amsthm}
\renewcommand{\Im}{\mathop{\mathrm{Im}}\nolimits}
\renewcommand{\Re}{\mathop{\mathrm{Re}}\nolimits}

\begin{document}
\title{Generation of Non-Gaussian States in the Squeezed State Entanglement Scheme}
\author{E. N. Bashmakova$^{1}$} 
\author{S. B. Korolev$^{1,2}$} 
\author{T. Yu. Golubeva$^{1}$}
\address{${}^1$ St. \,Petersburg \,State \,University, \,Universitetskaya \,nab. \,7/9, \,St. \,Petersburg, 199034, Russia}
\affiliation{${}^2$ Laboratory \,of \,Quantum \,Engineering \,of \,Light, \,South \,Ural \,State \,University, \,pr. \,Lenina \,76, \,Chelyabinsk, 454080, Russia}

\begin{abstract}
The paper considers the possibility of generating different non-Gaussian states using the entangled state photon measurement scheme. In the paper, we have proposed a way to explicitly find the wave function and the Wigner function of the output state of this scheme. Moreover, the solutions found are not restricted to any particular case, but have maximum generality (depend on the number of measured photons and on all parameters of the scheme). Such a notation allowed us to carry out a complete analysis of the output states, depending on the scheme parameters. Using explicit expressions, we have analyzed the magnitude of non-Gaussianity of the output states, and we have revealed which particular states can be obtained in the proposed scheme. We have considered in detail a particular case of measurement (single photon measurement) and have shown that using explicit expressions for the output state wave function one can find scheme parameters to obtain states suitable for quantum error correction codes with a large fidelity value and high probability. The Schrodinger’s cat state with amplitude $\alpha=2$ can be obtained with fidelity $F\approx 0.88$ and probability 18 percent, and the squeezed Schrodinger’s cat state ($\alpha=0.5$, $R=1$) with fidelity $F\approx 0.98$ and probability 22 percent.
\end{abstract}

\maketitle

\section{INTRODUCTION}
Non-Gaussian quantum states are an essential part of continuous variable quantum computation technique. The use of continuous variables allows one to build deterministic schemes that give a significant measurement result each time they are addressed. To achieve the universality of quantum computation, with continuous variables, it is necessary to be able to implement at least one non-Gaussian  transformation  \cite{Braunstein_2005,Lloyd_1999}. That is, it  makes non-Gaussian operations and states essential elements of quantum computing in continuous variables. The second reason for the tremendous interest in non-Gaussian states is the potential possibility of using such states in the error correction procedure. According to the No-Go theorem \cite{Niset2009} Gaussian states cannot be used to correct errors in Gaussian operations. As a result, non-Gaussian states are required for performing fault-tolerant universal quantum computations. 

Non-Gaussian quantum states are also considered as a potential resource for quantum cryptography and quantum communication \cite{Lee2019,Guo2019}. In addition to the entire above, non-Gaussian states can be used to improve existing protocols for quantum optics and quantum information. For example, in the problem of the continuous variable quantum teleportation protocol, using auxiliary non-Gaussian states, one can significantly reduce  the computation error \cite{Asavanant2021,Zinatullin2023,Zinatullin2021}. However, to date the methods for generating non-Gaussian states with the required properties have been developed to turn out to be very difficult for the experimental implementation \cite{Sychev2017,Etesse2015}. The class of non-Gaussian states involved in quantum information protocols is very narrow and limited by the stringent requirements \cite{Ralph_2003,Hastrup_2022}.

 Among non-Gaussian quantum states, of particular interest are the so-called Schrodinger cat states \cite{Sychev2017,Buzek1995}, which are a superposition of two coherent states $|\alpha\rangle$ and $|-\alpha\rangle$. The increased interest in such states is due to the prospect of their use as a resource for quantum error correction codes. However, it has been shown in \cite{Ralph_2003,Hastrup_2022} that the amplitude $\alpha = 2$  is the minimum required for effective operation of quantum error correction protocols. Nowadays, many protocols of Schrodinger’s cat generation \cite{Sychev2017,Ourjoumtsev_cat2007, Polzik2006, Huang2015, Ulanov2016, Gerrits2010, Takahashi2008, Baeva2022,Podoshvedov_2023,Thekkadath2020} have been proposed. At the same time, the experimentally achievable in the optical range values are $|\alpha| \leq 1.9$  \cite{Sychev2017, Ourjoumtsev_cat2007, Huang2015, Ulanov2016, Gerrits2010, Takahashi2008}. Such amplitude values are insufficient for the operation of error correction protocols.
 
 The standard method for Schrodinger’s cats states generating is the photon subtraction method. In it, a squeezed vacuum falls on a beam splitter with a low reflectance coefficient, the second input of which is not illuminated. The states in one of the channels are detected by photon number resolving detector (PNRD). However, to achieve $|\alpha| \geq 2$  values the number of detected photons should be $n \geq 4$ \cite{Dakna1997}. In the work \cite{Takase2021} it has been proposed a generalized photon subtraction method, where two orthogonal squeezed vacuum states are mixed at a beam splitter. By detecting photons in one of channels heralds the generation of Schrodinger's cat states in the other one.

Another interesting type of non-Gaussian quantum states is so-called the squeezed Schrodinger's cat state \cite{Grimsmo2020}. Based on them, the code of correction of both phase and photon-loss errors has been developed. Unlike the traditional Schrodinger's cats states, the squeezed Schrodinger's cats states used for the error correction should have a large squeezing degree and a small amplitude $\alpha$. Therefore, to build the error correction code using the squeezed Schrodinger's cat states, there are lower experimental requirements for the amplitude $|\alpha|$.

In this paper, we consider a scheme for generating quantum non-Gaussian states in which two squeezed vacuum states are mixed at the beam splitter, and then the PNRD detects photons in one channel. Unlike the authors of \cite{Takase2021}, we analyze the non-Gaussianity and study its dependence on the scheme parameters and entanglement degree. The main feature of our work is that we were able to obtain the analytical description of the entire protocol for an arbitrary number of detected photons. The analytical expressions allowed us to identify the most important parameters affecting the non-Gaussianity of the output states. As a measure of non-Gaussianity, we estimate the Wigner function negativity \cite{Kenfack2004}. We demonstrate that our approach for generating various non-Gaussian states with high fidelity and high probability allows us to reduce the requirements for the number of detected photons.

\section{Analytical description of non-Gaussian state generation}
\begin{figure}[H]
    \centering
    \includegraphics[scale=1]{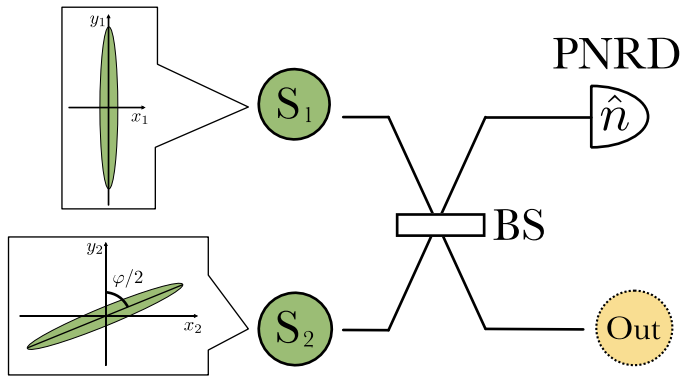}
    \caption{Scheme for generating non-Gaussian states.  In the diagram: $S_1$ and $S_2$ are two squeezed vacuum states, BS is a beam splitter, and PNRD is a photon number resolving detector.}
    \label{fig:SVS}
\end{figure}

To generate a non-Gaussian state, we will consider the scheme shown in Fig. \ref{fig:SVS}.
In the scheme, two squeezed vacuum states are mixed at the beam splitter (BS). The amplitude transmission and reflection coefficients of the beam splitter are $t$ and $\rho$, respectively. One measures the state at one of the beam splitter outputs using a photon number resolving detector (PNRD). The unmeasured state at the other output we will call the output state (Out state). In \cite{Takase2021}, the authors have already considered a similar scheme for the possibility of generating Schrodinger cat states. In our work, we want to make a more comprehensive analysis of this scheme. We want to investigate the magnitude of the non-Gaussianity of the output state obtained in this scheme. To understand how this magnitude dependent on the scheme parameters, as well as on the entanglement degree of quantum oscillators. For such an analysis, we need to obtain an explicit expression of the wave function and the Wigner function of the output state. Let us move on to finding these functions.

To understand what the output state will be, let us look at each step of the scheme in detail. We begin by describing the squeezed vacuum states that are mixed at the beam splitter. In the general case, the squeezed vacuum state (SVS) is determined by two parameters: the degree of squeezing $r$ and the rotation angle on the phase plane $\varphi/2$. One can define such a squeezed state in the coordinate representation as follows:
\begin{align} 
    |\text{SVS};r,\varphi\rangle =\int  dx \Psi \left(x;r,\varphi\right)|x\rangle ,
\end{align}
where the wave function of an arbitrary squeezed vacuum state is introduced:
\begin{align} \label{sq_s}
    \Psi(x;r,\varphi)=\frac{1}{ \sqrt[^4]{\pi}\left(\cosh{2r}-\cos{\varphi}\sinh{2r}\right)^{1/4}}\exp \left(-\frac{x^2}{2}\frac{1+i\sin{\varphi}\sinh{2r}}{\cosh{2r}-\cos{\varphi}\sinh{2r}}\right).
\end{align}
The choice of the angle $\varphi$ corresponds to the rotation of the state on the phase plane through the angle $\varphi/2$. That is, when the relative phase is $\varphi=\pi$, the states with the same squeezing degree $r$ will be squeezed orthogonally.

In the problem under consideration, we do not care about the common phase of the two squeezed states. We will assume that the first squeezed oscillator has a phase equal to zero, while the phase of the second oscillator is equal to $\varphi$ ($\varphi \in \left[0,\pi\right]$). In addition, for simplicity, we will assume that the degree of squeezing of the two oscillators is the same. Considering all this, we can write the state of the system before the beam splitter in the following form:
\begin{align}
    |\text{SVS};r,0\rangle _1 |\text{SVS};r,\varphi\rangle_2 =\int \int  dx_1 dx_2 \Psi \left(x_1;r,0\right)\Psi \left(x_2;r,\varphi\right)|x_1\rangle _1 |x_2\rangle _2.
\end{align}

These two oscillators then interfere on a beam splitter, which transforms them as follows:
\begin{align}
    \hat{BS}_{12} |\text{SVS};r,0\rangle _1 |\text{SVS};r,\varphi\rangle_2 =\int \int  dx_1 dx_2 \Psi \left(t x_1+\rho x_2;r,0\right)\Psi \left(-\rho x_1+t x_2;r,\varphi\right)|x_1\rangle _1 |x_2\rangle _2.
\end{align}
Here $t$ and $\rho=\sqrt{1-t^2}$ are real amplitude transmission and reflection coefficients, respectively.

Next, the state in the first channel is measured on the PNRD. Let us assume that the result of the measurement is the number $n$. Such a measurement leads to the following non-normalized state vector: 
\begin{align}
    {}_1\langle n|\hat{BS}_{12} |\text{SVS};r,0\rangle _1 |\text{SVS};r,\varphi\rangle_2=\int \int dx_1 dx_2 \Psi \left(t x_1+\rho x_2;r,0\right)\Psi \left(-\rho x_1+tx_2;r,\varphi\right){}_1\langle n|x_1\rangle _1 |x_2\rangle _2. 
\end{align}
Given that $${}_1\langle n|x_1\rangle _1=\frac{1}{\sqrt[^4]{\pi}2^{n/2}\sqrt{n!}}e^{-x_1^2/2}H_n(x_1) ,$$ as well as an explicit expression for the wave functions of the squeezed state (\ref{sq_s}), we can write the wave function of the output state as:
\begin{align}  \label{int_out}
  \Psi_{out}(x_2) =\frac{e^{-cx_2^2}}{\sqrt{N}} \int e^{-ax_1^2+bx_1x_2}H_n(x_1).
\end{align}
Here $N=\int|\Psi_{out}(x_2)|^2 dx_2$ is the normalization of the output state wave function, $H_n(x)$ is the Hermite polynomial, and the parameters are given by the following expressions: 
\begin{align}
   &a=\frac{1}{2} \left(1+e^{2 r} t^2-\frac{\left(t^2-1\right) (1+i \sin \varphi  \sinh 2r)}{\cosh 2 r-\cos \varphi \sinh 2 r}\right), \label{param_1}\\
    &b=\frac{\left(e^{4 r}-1\right) t \sqrt{1-t^2} \sin \varphi/2}{e^{2 r} \sin \varphi/2+i \cos \varphi/2},\\
    &c=\frac{1}{2} \left(\frac{t^2 (1+i \sin \varphi \sinh 2 r)}{\cosh 2 r-\cos \varphi  \sinh 2 r}-e^{2 r} \left(t^2-1\right)\right) . \label{param_3}
\end{align}

To study the output state, it will be convenient for us to explicitly calculate the integral in the Eq. (\ref{int_out}). To do this, let us use the generalized Hermite polynomial $H_n(x,y)$, which is defined by the following series:
\begin{align}
    H_n(x,y)=n!\sum \limits_{k=0}^{\lfloor \frac{n}{2}\rfloor} \frac{x^{n-2k}y^k}{(n-2k)!k!}.
    \end{align}
Here $\lfloor \frac{n}{2}\rfloor$ is the floor function. In \cite{Babusci2012} it is shown that using the generalized Hermite polynomial, one can write an analytical expression for an integral of the form:  
\begin{align} \label{an_int}
    \int dx H_n(lx+e,f)e^{-\alpha x^2+\beta x}=\sqrt{\frac{\pi}{\alpha}}\exp\left[\frac{\beta^2}{4\alpha}\right]H_n\left(e+\frac{\beta l}{2\alpha},f+\frac{l^2}{4\alpha}\right) .
\end{align}
Given the Eq. (\ref{an_int}), as well as the relationship between the ordinary and generalized Hermite polynomials
\begin{align}
    H_n(x)=H_n(2x,-1),
\end{align}
 we can write an analytical expression for the output state in the form:
\begin{align} \label{out_an}
  \Psi_{out}(x_2) =\frac{e^{-cx_2^2}}{\sqrt{N}} \int e^{-ax_1^2+bx_1x_2}H_n(2x_1,-1) =\sqrt{\frac{\pi}{aN}} \exp\left[\left(\frac{b^2}{4a}-c\right)x_2^2\right]H_n\left(\frac{bx_2}{a},\frac{1-a}{a}\right),
\end{align}
where $a$, $b$ and $c$ are defined by Eqs. (\ref{param_1})-(\ref{param_3}).

Knowing the analytic expression for the wave function, we can find an analytic expression for the normalization of this function. To do this, we need to calculate an integral of the form: 
\begin{align}
 N=  \int|\Psi_{out}(x_2)|^2 dx_2= \frac{\pi}{|a|}\int dx_2 \exp\left[\left(\frac{b^2}{4a}-c+\frac{b^{*2}}{4a^*}-c^*\right)x_2^2\right] H_n\left(\frac{bx_2}{a},\frac{1-a}{a}\right)H_n\left(\frac{b^*x_2}{a^*},\frac{1-a^*}{a^*}\right).
\end{align}
This integral has the following representation:
\begin{align}
    N=\frac{\pi}{|a|} I_{n,n}\left(\frac{b}{a},0,\frac{1-a}{a},\frac{d^*}{a^*},0,\frac{1-a^*}{a^*},\left(\frac{b^2}{4a}-c+\frac{b^{*2}}{4a^*}-c^*\right),0\right).
\end{align}
Here we have introduced an integral of the form:
 \begin{align}
  I_{m,n}\left(d_1,e_1,f_1,d_2,e_2,f_2,\alpha,\beta\right)=\int dx H_m(d_1x+e_1,f_1)H_n(d_2x+e_2,f_2)e^{-\alpha x^2+\beta x}, 
 \end{align}
which can be written analytically using the two-index Hermite polynomial \cite{DATTOLI2000111,Babusci2012}: 
  \begin{align}
     I_{m,n}\left(d_1,e_1,f_1,d_2,e_2,f_2,\alpha,\beta\right)= \sqrt{\frac{\pi}{\alpha}}\exp \left[\frac{\beta^2}{4\alpha}\right]H_{m,n}\left(e_1+\frac{d_1}{2\alpha}\beta,f_1+\frac{d_1^2}{4\alpha},e_2+\frac{d_2}{2\alpha}\beta,f_2+\frac{d_2^2}{4\alpha}|\frac{d_1d_2}{2\alpha}\right).
 \end{align}
The two-index Hermite polynomial is given by the following expression:
 \begin{align} \label{tiHp}
     H_{m,n}\left(x,y,w,z|t\right)=\sum\limits _{k=0}^{\min \left(m,n\right)}\frac{n!m!}{(m-k)!(n-k)!k!}t^kH_{m-k}(x,y)H_{n-k}(w,z).
 \end{align}

Having simplified all the expressions, we can write the final form of the output state wave function:
\begin{multline} \label{vfout}
    \Psi_{out}(x_2) =\frac{1}{\sqrt{N(n,r,t,\varphi)}}\sqrt{\frac{e^r \left(e^{2 r}+i \cot \frac{\varphi }{2}\right)}{\xi}}\exp \left[-\frac{x_2^2 \left(e^r-\left(1-e^{i \varphi }\right) t^2 \sinh
   r\right)}{2 \left(e^{-r}+\left(1-e^{i \varphi }\right) t^2 \sinh r\right)}\right]\\
   \times H_n\left(-\frac{2 e^{r} \gamma }{\xi }x_2,\tanh r \left(\frac{2(1-t^2)\sin \frac{\varphi}{2}}{\xi}-1\right)\right).
\end{multline}
and its normalizations:
\begin{align}
    N\left(n,r,t,\varphi\right)=\sqrt{\pi } n! \left(\frac{2\gamma^2}{\gamma^2 +1}\right)^n \sqrt{\frac{\cot
   ^2\frac{\varphi }{2}+e^{4 r}}{\gamma^2+1}} \, _2F_1\left(\frac{1-n}{2},-\frac{n}{2};1;\frac{
   \sinh ^2r-\gamma^2}{\gamma^4\cosh^2 r}\right),
\end{align}
where, for simplicity, we introduced notations:
\begin{align}
& \xi = \left(1+\left(e^{2 r}-1\right) t^2\right)\sin \frac{\varphi }{2} +i \cos \frac{\varphi }{2} ,\\
&\gamma  =  2 t \sqrt{1-t^2} \sin \frac{\varphi }{2} \sinh r \label{eq_gamma},
\end{align}
and $_2F_1(x,y;k,z)$ is the hypergeometric function. Given the explicit form of the wave function, we can write an explicit probability function to obtain a given state:
\begin{align} \label{Prob}
    P\left(n,r,t,\varphi\right)=\frac{\sin \frac{\varphi}{2} }{2^{n}n! e^{r} \cosh r  \sqrt{\pi \left(\cosh 2 r-\cos \varphi \sinh 2 r\right)}}N\left(n,r,t,\varphi\right) .
    \end{align}

For further analysis of the properties of the output state, we also need its Wigner function, which has the following analytical expression:
\begin{multline}
 W_n\left(x,p\right)=\frac{1}{N\left(n,r,t,\varphi\right)}\sqrt{\frac{\cot^2 \frac{\varphi}{2}+e^{4r}}{\pi\left(\gamma^2+1\right)}}  \exp \left[-\frac {p^2\frac{|\xi|^2}{e^{2r}}+x^2\frac{ \left(\gamma ^2+1\right)^2+\eta^ 2}{e^{-2r}|\xi|^2}+2xp\eta}{\gamma
   ^2+1}\right]\\
   \times\sum \limits _{k=0}^{n} \binom{n}{k}^2k!\left(-\frac{2 \gamma ^2}{\gamma ^2+1}\right)^k\left|H_{n-k}\left(-\frac{2 \gamma}{\gamma ^2+1}\left(\frac{\xi -2 t^2 \sin \frac{\varphi}{2}\sinh 2 r}{e^{-r}}x+i\frac{\xi}{e^r}p\right),-\frac{t^2+e^{i \varphi } \left(1-t^2\right)}{\left(\gamma ^2+1\right)\coth r}\right)\right|^2,
\end{multline}
where, for ease of notation, we introduced the following notation:
\begin{align}
   \eta= t^2\sin \varphi \sinh 2 r,
\end{align}
Appendix \ref{sec_WF} shows the derivation of the output state's Wigner function. 

Summarizing this section, it can be noted that using the generalized Hermite polynomial, one can write both the wave function and the Wigner function of the output state in the most general form without any approximations. Obtained analytical solutions will help us to fully explore the properties of the output states.

\section{STUDY OF QUANTUM NON-GAUSSIANITY OUTPUT STATE}
\subsection{Wigner negativity}
As mentioned earlier, it is possible to generate non-Gaussian states in the scheme we have presented. In the scheme non-Gaussianity is induced by the non-Gaussian measurement (measurement of the number of photons) of the Gaussian entangled state. Due to the entanglement between the oscillators, non-Gaussian measurements of one state lead to the teleportation of non-Gaussianity to an unmeasured state. Thus, our challenge is to perform this teleportation in the most effective way, as well as to understand how this transfer can be controlled. Let us estimate the non-Gaussianity depending on the scheme parameters. Let us bring out the quantum states obtained at certain parameters.

We will estimate the magnitude of the non-Gaussianity using the Wigner negativity, which is determined through the Wigner function as follows \cite{Kenfack_2004}:
\begin{align}
    \mathcal{N}_n\left(r,t,\varphi\right)=\int \left|W_n\left(x,p\right)\right|dxdp-1,
\end{align}
where the dependence of the Wigner function on the scheme parameters is omitted for brevity. The negativity of the Gaussian state is equal to zero. Non-Gaussian states have non-zero Wigner negativity \cite{Francesco_2018}. The greater the magnitude of the state’s non-Gaussianity, the higher the magnitude of its negativity.

In the scheme, there are several parameters that we can vary. Firstly, the parameters of the initial squeezed states: the squeezing degree as well as the value of the relative phase. Secondly, it is the beam splitter transmission coefficient. In addition, the obtained non-Gaussian state will strongly depend on the number of detected photons.

\begin{figure}[H]
    \centering
    \includegraphics[scale=0.9]{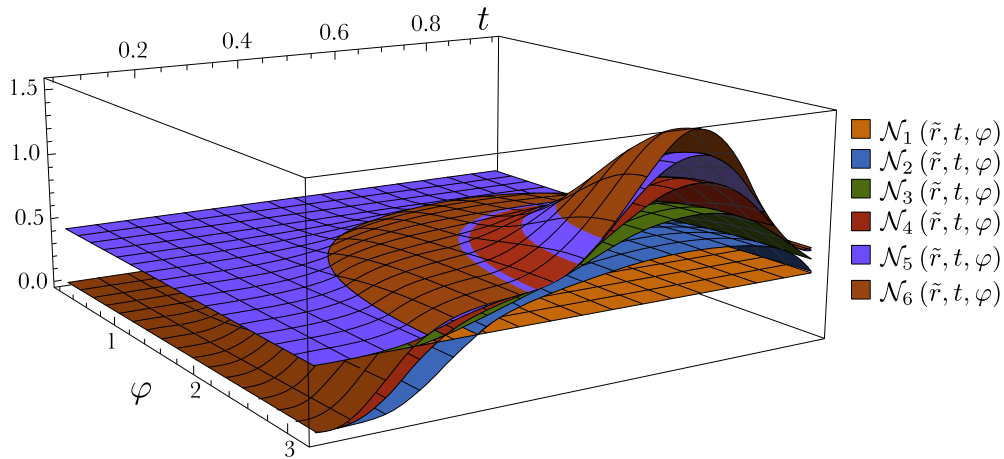}
    \caption{Dependence of the Wigner negativity on the value of the relative phase of the squeezed states and on the beam splitter transmission coefficient at a certain value of the number of detected photons. The squeezing degree of all states is identical and equal to $\tilde{r}=8\ln 10/20$, which corresponds to -$8$ dB.}
    \label{fig:neg_all_surf}
\end{figure}

Let us first study the non-Gaussianity of the output state depending on the value of the relative phase of the two squeezed states, on the beam splitter transmission coefficient and on the number of detected photons.  We will consider the squeezing degree of two oscillators as $-8$ dB. The Fig. \ref{fig:neg_all_surf} shows the Wigner negativity at a fixed squeezing degree  depending on the other parameters. The surfaces of different colours correspond to measurements of different numbers of photons in the first channel.

By looking at the figure several patterns can be identified. Firstly, we see that all the surfaces are divided into two types depending on the parity of the number of detected photons. If the number of detected photons is even, the negativity of the output states will decrease to zero at the edges of the region. If an odd number of photons has been measured in the scheme, then the negativity of the output state will always exceed $0.426$ (which corresponds to the magnitude of negativity of the first Fock state). In addition, for each surface, it is possible to determine regions of parameters at which the negativity does not increase with the number of detected photons. Let us define such areas as a "negativity plateau"\,. One can also identify the areas of parameters in which there is a significant increase of negativity from the number of detected photons (let's define as a "mountain of negativity"). Let us consider each of these areas in more detail and justify, from a physical point of view, the negativity in these regions.

\subsection{The entanglement degree of output state}
\begin{figure}[H]
    \centering
    \includegraphics[scale=0.75]{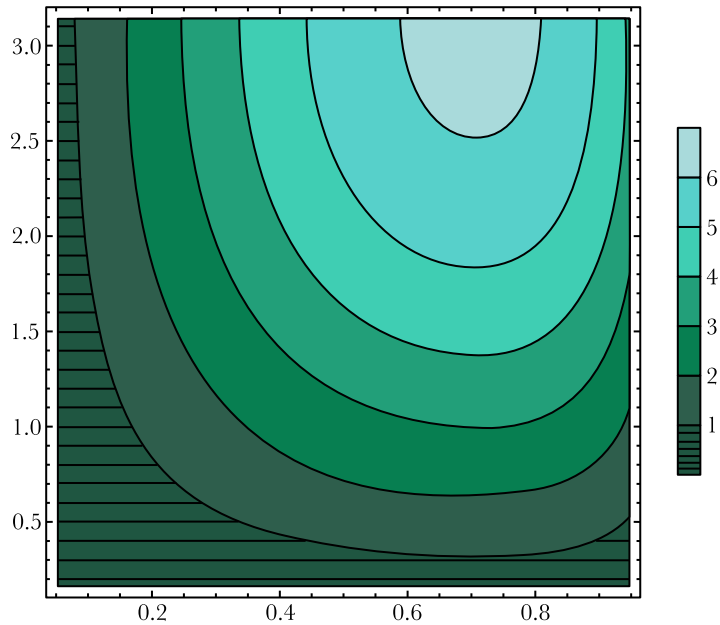}
    \caption{The entanglement degree of the output state depending on the transmission coefficient $t$ and the relative phase of the squeezed oscillators $\varphi$. The squeezing degree of two oscillators is equal to $-8$ dB.}
    \label{fig:entangl}
\end{figure}

To fully understand the values of the Wigner negativity for certain parameters $\varphi$ and $t$, let us estimate the entanglement degree for this system. To estimate the entanglement, we will use the van Loock–Furusawa criterion \cite{Furusawa_2003}. Using this criterion, we can write an inequality on the squeezed parameter of the initial oscillators:
\begin{align} \label{vLFc}
    \frac{ 2t\sqrt{1-t^2}\sin \frac{\varphi}{2}}{e^{-2r}}>1.
\end{align}
Appendix \ref{sec_EC} shows the derivation of this inequality. Based on the presented criterion, it follows that the output state will be entangled if, for the specific beam splitter transmission coefficient and the relative phase of the squeezed states, one chooses the squeezing degree so that condition (\ref{vLFc}) is satisfied. Moreover, the greater the value on the left-hand side, the greater the entanglement will be. According to the obtained inequality, let us identify the region of the parameters $\varphi$ and $t$ at which the system will be entangled. As before, we will consider the initial squeezing degree equal to $-8$ dB. The contour plot of the entanglement degree of output states is shown in Fig. \ref{fig:entangl}. 
The graph shows that for any value of the parameters of the shaded area and the squeezing degree of the oscillators at $-8$ dB, the output state will be separable. For the other parameters, the state will be entangled. The entanglement degree increases to the point $\left(t=\frac{1}{\sqrt{2}},\varphi=\pi\right)$, where the maximum value is reached. In other words,  the output state will  be as entangled as possible if the scheme contains a symmetrical 50:50 beam splitter and the initial oscillators squeezed orthogonally.

\subsection{Impact of entanglement degree on the non-Gaussianity of the output state}

\begin{figure}[H]
    \centering
    \includegraphics[scale=0.63]{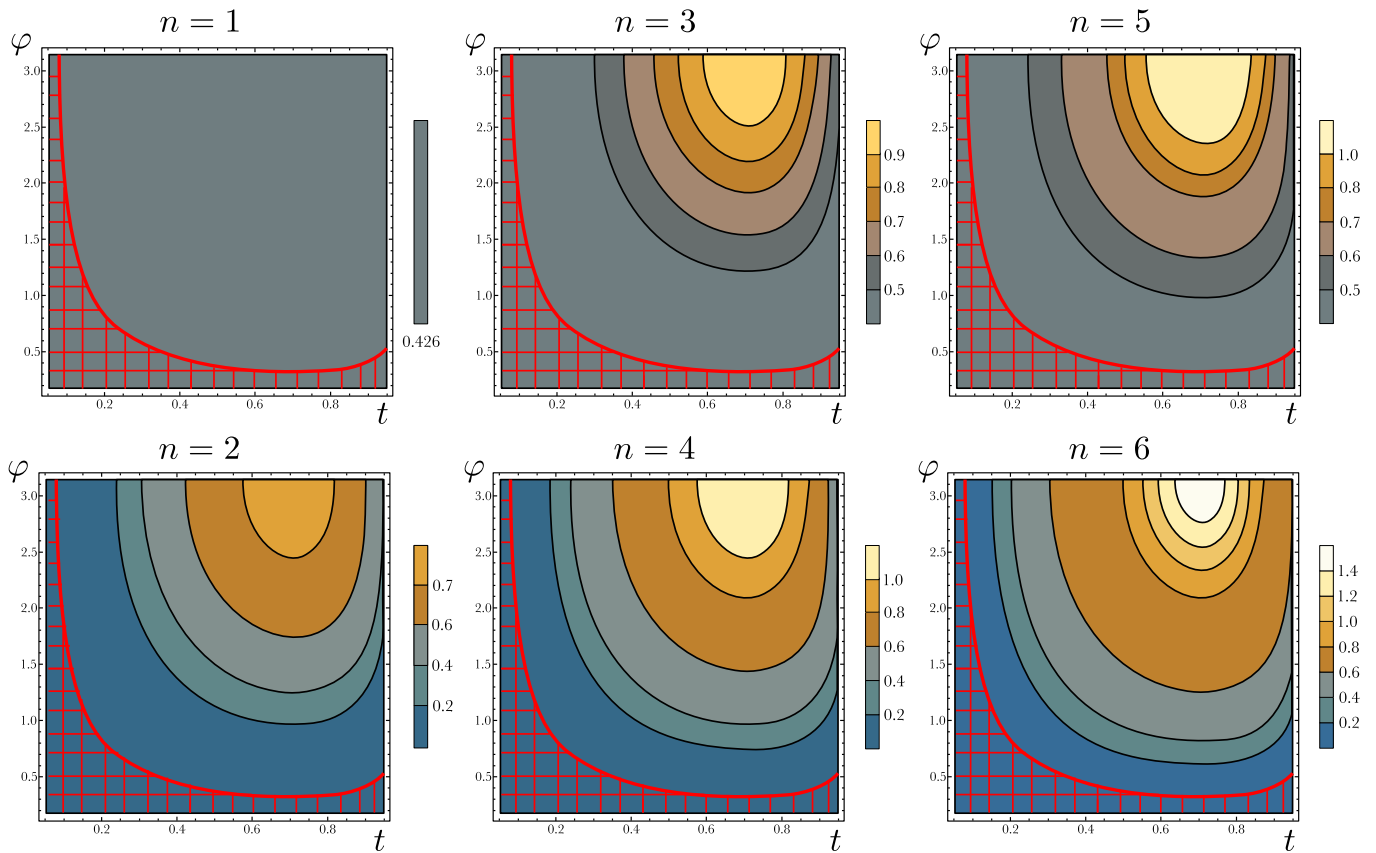}
    \caption{A graph of the dependence of Wigner negativity on the relative phase of the squeezed oscillators $\varphi$ and transmission coefficient $t$. On the graph: the red checkered area indicates the parameters for which the output state will not be entangled. Different frames correspond to the negativities of output states at the different number of detected photons.}
    \label{fig:ent_neg}
\end{figure}

Let us compare the Wigner negativity landscape obtained above (Fig. \ref{fig:neg_all_surf}) with the areas identified by the van Loock–Furusawa criterion (Fig. \ref{fig:entangl}). To do it, let us show on a single contour plot the negativity surfaces, and the region of the parameters for which the output state will not be entangled. The surface data is shown in Fig. \ref{fig:ent_neg}.

Comparing Fig. \ref{fig:ent_neg} and \ref{fig:neg_all_surf} one can see that the shape of the boundaries of the different negativities coincide exactly with the boundaries of the entanglement degree. It indicates a direct relationship between the magnitude of non-Gaussianity and the entanglement degree.

The graphs show that as the entanglement degree increases, so does the non-Gaussianity reaching its maximum value at the point with maximum entanglement degree ($\varphi=\pi$ and $t=\frac{1}{\sqrt{2}}$). As the above, the point corresponds to the case where two orthogonally squeezed states are mixed at a symmetrical beam splitter. As a result, the output state is a two-mode squeezed state, which can be decomposed into Fock states as follows: $|TMSS\rangle _{12} =\sum a_n|n\rangle_1|n\rangle_2$. This means that the maximum non-Gaussianity of the output state at $n$ detected photons will correspond to the non-Gaussianity of the $n$-th Fock state.

Let's justify the magnitude of the negativity in the region of parameters where the oscillators are not entangled. As we can see from Figs. \ref{fig:neg_all_surf} and \ref{fig:ent_neg} there can be two potential magnitudes of negativity in this region, depending on the parity of the number of detected photons. In addition, the magnitude of the negativity in the region of parameters under consideration does not increase with the number of detected photons.

Since the oscillators are not entangled, there are the independent Gaussian states at the output modes of the beam splitter. However, as the initial squeezed states are even (the squeezed states are only decomposed into even Fock states), the output states are also even. Any odd number of detected photons $n$ will be the sum of an even number of photons of a $2n_i$ state and an odd number of another $2n_d+1$ states. The most probable case (for this region of parameters) is that exactly single photon is detected. As a result,  there are a photon subtracted or added sates at the output mode. It is necessary to mention that according to \cite{Francesco_2018}, such states have the same negativity equal to $0.426$. This is exactly the value we see the region without entanglement for the case of an odd number of detected photons.

If an even number of photons have been detected, it indicates that the PNRD detects an even number of photons. At the same time for this range of parameters, an even number can only be achieved when there are zero photons from the second initial oscillator. In other words, when there has been no  single-photon subtraction from the second state the out state remains Gaussian. The negativity of such a state is equal to zero in this case, as it is shown in the presented graphs.

Consider the last interesting example of a single photon being detected. As follows from Fig. \ref{fig:ent_neg} the Wigner negativity in this case is constant and does not increase with growth of entanglement. As we have already investigated, the minimum magnitude of negativity for an odd number of detected photons is $0.426$. The maximal Wigner negativity in that case is equal to the negativity of the one-photon Fock state, which is also $0.462$. Thus, in the case of single-photon detection the Wigner negativity will be constant independent of the selected parameters.

\subsection{Effect of the squeezing degree on non-Gaussianity of the output state}
\begin{figure}[H]
    \centering
    \includegraphics[scale=0.95]{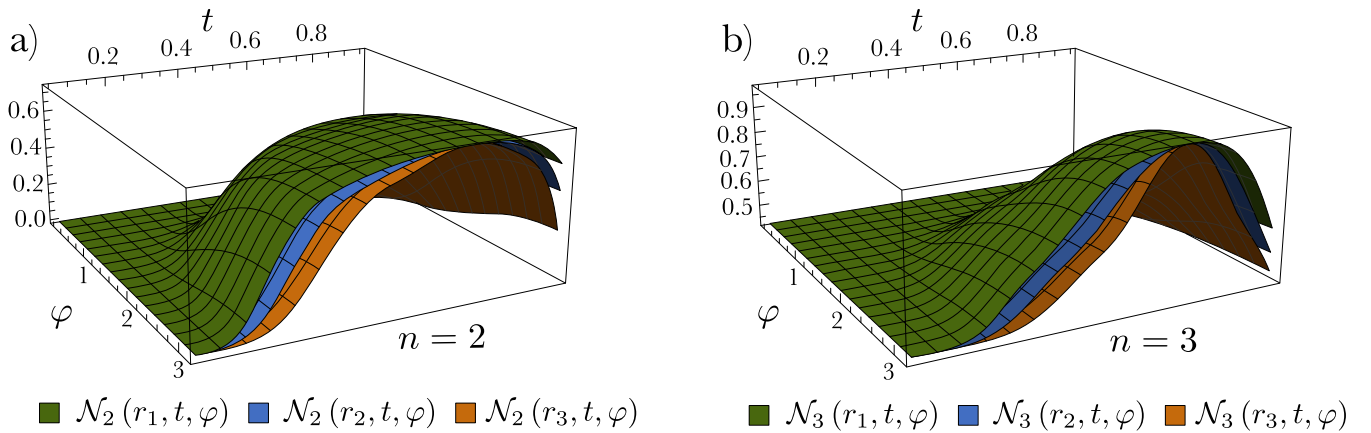}
    \caption{A graph of the dependence of the Wigner negativity on the relative phase of the squeezed states and the beam splitter transmission coefficient at the different squeezing degree. The graphs show two cases: a) detection of two photons, b) detection of three photons. The squeezing degree equals: $r_1={10\ln10}/{20}$ (10 dB), $r_2={8\ln10}/{20}$ (8 dB), $r_3={6\ln10}/{20}$ (6 dB). }
    \label{fig:sq_neg}
\end{figure}

We have already considered the influence of the beam splitter transmission coefficient and relative phase of the squeezed oscillators as well as  the number of detected photons on the non-Gaussianity of the output state. Now let us investigate the dependence of the non-Gaussianity on the squeezing degree of the oscillators. To do this, we plot the dependence of the Wigner negativity on the parameters $\varphi$ and $t$ for a certain number of detected photons and different squeezing degree. These dependencies are shown in Fig. \ref{fig:sq_neg}.
The graphs demonstrate that as the squeezing degree increases, the Wigner negativity increases at all points in which there is the observed entanglement. This is due to the squeezing degree $r$ increases, so does the entanglement degree, which directly follows from the inequality (\ref{vLFc}). At the point of the maximum entanglement (the symmetrical beam splitter and orthogonally squeezed states), there is no dependence of the magnitude of the Wigner negativity on the squeezing degree. The result is that at the point of the maximum entanglement one has a Fock state, which the negativity depends only on the number of the detected photons.

\section{Quantum states implemented in the presented scheme}
\subsection{Variety of quantum states generated in the scheme}
\begin{figure}[H]
    \centering
    \includegraphics[scale=0.66]{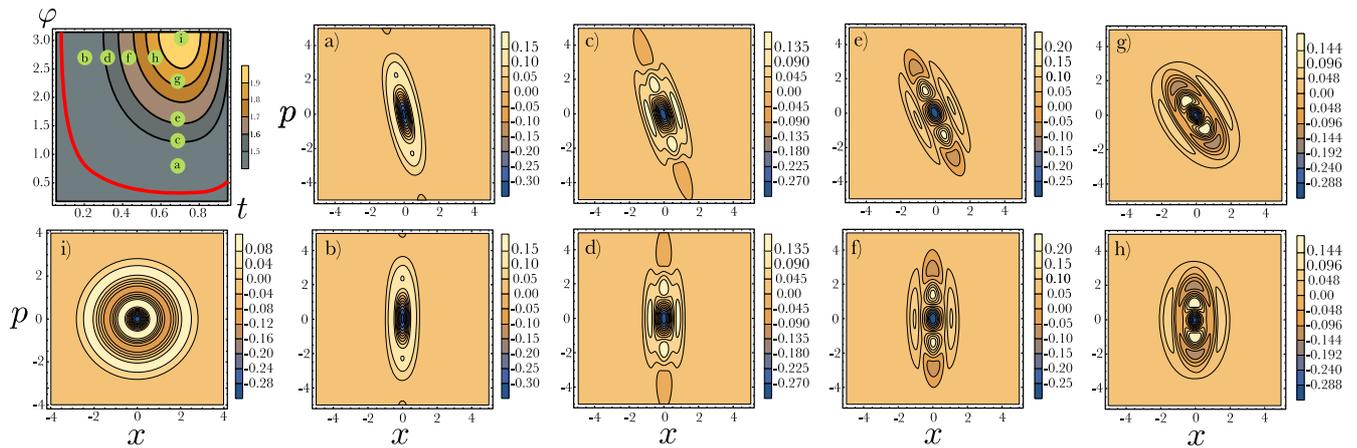}
    \caption{Wigner functions of the output state depending on the magnitude of negativity. In the upper left corner, a graph of the dependence of negativity on the angle $\varphi$ and the transmission coefficient $t$ for the case of measuring 3 photons and squeezing of $-8$ dB is shown. On this graph, the points a)-i) denote the parameters used in the construction of the Wigner functions of the output states depicted in frames a)-i), respectively.}
    \label{fig:WanL}
\end{figure}

After we have found out which parameters affect the non-Gaussianity of the output state, we can proceed to the study of the states that are obtained in the considered scheme. To understand the type of states that can be implemented, let us plot the Wigner functions of the states with certain negativity values. Fig. \ref{fig:WanL} shows this graph.

The graphs show the Wigner functions of pairs of states with the same negativity value. All presented states are obtained by measuring three photons and squeezing the original oscillators by $-8$ dB. The graphs show that the states with the same negativity value differ from each other only by a rotation on the phase plane. Moreover, the type of state depends only on the magnitude of negativity. The greater the magnitude of the negativity, the larger the region with a negative value of the Wigner function. 

In Fig. \ref{fig:WanL} one can observe three types of states: a) and b) are states similar to Schrodinger’s cat states; states e), f) are squeezed Schrodinger’s cat type states; i) Fock state. Ordinary and squeezed Schrödinger’s cat states are needed for coding quantum information in quantum error correction codes \cite{Grimsmo2020,Ralph_2003,Hastrup_2022,Schlegel_2022}. Therefore, let us examine our scheme for realizability of this type of states. Let us consider the feasibility of this or that of Schrodinger’s cat state for the most probable non-trivial outcome of measurements, when measuring a single photon.

\subsection{Schrodinger's cat state generation}
Let us consider the possibilities of our scheme to realize the Schrodinger’s cat state. For this purpose, let us consider the fidelity of the output state in the case of a single photon measurement with Schrodinger’s cat state. The following state vector defines the Schrodinger’s cat state:
\begin{align}
    |\text{cat}_{\pm}\rangle=\frac{1}{\sqrt{N_{cat_{\pm}}}}\left(|\alpha\rangle\pm |-\alpha\rangle\right),
\end{align}
where $N_{\pm}=2\left(1\pm e^{-2|\alpha|^2}\right)$ is the normalization factor. The state $|\text{cat}_{+}\rangle$ ($|\text{cat}_{-}\rangle$) is called even (odd), since it is decomposed into the even (odd) Fock states.

In this paper, we will be interested in the cat state with amplitude $\alpha=2$. As was shown in \cite{Ralph_2003,Hastrup_2022}, such amplitude is the minimum necessary for quantum error correction codes. 

As noted above, we consider the case when the PNRD has measured a single photon. Since the squeezed states are even, the detection of one photon in the measured channel means a change in the parity of the output state. The odd output state should be compared with the odd Schrodinger's cat state. The fidelity of the output state with measuring one photon and the odd Schrodinger’s cat state with amplitude $\alpha=2$ will have the following form:
\begin{align} \label{fid_cat}
    F_{\text{cat}_{-}}\left(r,t,\varphi\right)=\frac{4 \left(\gamma ^2+1\right)^{3/2} e^{-4 \tanh r \left(1-2t^2 \sin^2 \frac{\varphi}{2}\right)}}{\sinh 4 \cosh ^3r },
\end{align}
where, for simplicity, we use the previously introduced notation (\ref{eq_gamma}). Our goal is to optimize the scheme parameters so that the function (\ref{fid_cat}) takes the maximum value. The global maximum of the function will be reached with the following set of parameters:
\begin{align}
    &\varphi=\pi,\\
    &t= \tilde{t}\equiv\frac{1}{4} \sqrt{\left(\sqrt{73}-3\right) \coth r+8}.
\end{align}
However, this optimization is only valid if $r\geqslant \text{arccoth}\left(\frac{1}{8} \left(3+\sqrt{73}\right)\right))$, that is, if the squeezing degree is greater than $-7.42$ dB. In the optimal case, fidelity is equal to:
\begin{align}
     F_{\text{cat}_{-}}\left(r,\tilde{t},\pi\right)\approx 0.88.
\end{align}
We see that in the optimized case (with correctly chosen parameters $\varphi=\pi$ and $t=\tilde{t}$) the fidelity is a constant value and does not depend on the squeezing degree $r$. It turns out that for any orthogonally squeezed oscillators with a squeezing degree above a certain threshold value, it is possible to choose the beam splitter transmission coefficient so that the maximum value of the fidelity of the Schrödinger cat state with an amplitude $\alpha=2$ is approximately $88\%$.

Let us now estimate the probability of obtaining this state. To do this, we use the Eq. (\ref{Prob}) given that $n=1$, $\varphi=\pi$, and $t=\frac{1}{4} \sqrt{\left(\sqrt{73}-3\right) \coth r+8}$. The probability of getting an output state in this case is determined by the following expression: 
\begin{align}
    P(1,r,\tilde{t},\pi)=\frac{4 \sqrt{2} \left(32 \tanh ^2r+3 \sqrt{73}-41\right)}{\left(3\sqrt{73}-9\right)^{3/2}\cosh ^2r}.
\end{align}
The fact that the optimized fidelity is independent of the squeezing value means that the squeezing degree can be controlled to achieve, for example, a maximum probability. At the point $\cosh r={8}/{\sqrt{3 \sqrt{73}-9}}$ the probability function reaches a maximum value of about $0.18$.

Thus, we have obtained that having orthogonally squeezed oscillators with a squeezing degree of $-11.24$ dB, we can obtain Schrodinger's cat states (amplitude $\alpha=2$) with probability $18 \%$ and fidelity $88 \%$ if only one photon is measured. Let us now examine these same quantities when the generating a squeezed Schrodinger's cat state.


\subsection{Generation of squeezed Schrodinger's cat type states}
In this section, we will compare the output state got by measuring a single photon with the squeezed Schrodinger's cat state, which is given by the following expression:
\begin{align}
    |\text{Scat}_{\pm}\rangle=\frac{1}{\sqrt{N_{\text{Scat}_{\pm}}}}\left(|\alpha,R\rangle\pm|-\alpha,R\rangle\right), 
\end{align}
where $|\alpha,R\rangle=\hat{D}\left(\alpha\right)\hat{S}\left(R\right)|0\rangle$ is the squeezed coherent state, and $N_{\text{Scat}_{\pm}}$ is the normalization factor. As in the previous section, we will evaluate the possibility of generating a useful state, which can be applicable to quantum error correction codes. The authors of paper \cite{Schlegel_2022} proposed a correction code using the squeezed Schrodinger's cat states. In contrast to traditional codes using the ordinary Schrodinger's cat states, this code can correct both phase errors and photon loss errors. For this, it is important that the squeezed cat state has a large squeezing degree $R$ and a small amplitude $\alpha$. Depending on the photon loss rate and the magnitude of the phase errors, the ratio between the amplitude and squeezing degree of the cat is varied. In this paper, as an example of a squeezed Schrodinger's cat state, we will consider a state with $\alpha=0.5$, $R=1$. Such a state, according to the data of \cite{Schlegel_2022}, can protect information in channels with an average error rate.

Our primary goal in this subsection is to maximize the fidelity of the output state got by measuring one photon and the odd squeezed Schrodinger's cat state with $\alpha=0.5$ and $R=1$. The fidelity of these two states has the following form: 
\begin{align}
    F_{\text{Scat}_{-}}\left(r,t,\varphi\Big|R=1;\alpha=\frac{1}{2}\right)=\frac{2g\left(r,t,\varphi\right)^3e^{5}\left(\coth \frac{e^2}{4} -1\right)}{\sin \frac{\varphi}{2}} \exp
   \left(-\frac{e^4}{2} \Re\left[\frac{1}{1-\frac{2 e^r \cosh r}{1+\left(1-e^{i \varphi }\right) e^r t^2 \sinh r}-e^2}\right]\right),
\end{align}
where the following abbreviations have been introduced for simplicity of recording: 
\begin{align}
   &g\left(r,t,\varphi\right)=e^{r}\left| \frac{1+e^r \left(1-e^{i \varphi }\right) t^2 \sinh r}{e^r \left(e^2-1\right)
   \left(1-e^{i \varphi }\right)t^2 \sinh r+e^{2 r}+e^2}\right|\sqrt{ \frac{1+4 t^2
   \left(1-t^2\right) \sinh ^2r \sin^2 \frac{\varphi}{2}}{\cot ^2\frac{\varphi }{2}+\left(\left(e^{2 r}-1\right) t^2+1\right)^2}}.
\end{align}
The presented function takes the maximum value with the following parameters:
\begin{align}
    &\varphi=\pi,\\
    &t=\tau\equiv\sqrt{\frac{\left(e^2 \sqrt{36+e^4}-3\left(1+ e^4\right)\right) \coth r+4 e^4-3}{8 e^4-6}},\\
    &r \geqslant \text{arccoth}\left(\frac{3+3 e^4+e^2 \sqrt{36+e^4}}{2 e^4-3}\right)\approx 0.48.
\end{align}
It turns out that for any value of squeezing degree, which is greater than a certain value, one can choose the transmission coefficient so that the maximum fidelity will be approximately equal to
\begin{align}
    F_{\text{Scat}_{-}}\left(r,\tau,\varphi\Big|R=1;\alpha=\frac{1}{2}\right) \approx 0.98
\end{align}
Since this is true for any squeezing degree, the value $r$ can be used to increase the probability of obtaining the state. The following formula gives this probability: 
\begin{align}
  P(1,r,\tau,\pi)=\frac{6 e^2 \left(4 e^4-3\right) \left(e^6-13 e^2+\sqrt{36+e^4}\left(e^4+1 \right)\right) \cosh^2r-\left(4 e^4-3\right)^3 }{6 \sqrt{6}
   e^3 \left(e^6-13 e^2+\sqrt{36+e^4}\left(e^4+1 \right)\right)^{3/2}\cosh^4r}.
\end{align}
This function takes the maximum value $P(1,r,\tau,\pi)\approx 0.22$ at $r\approx 1.032$ ($8.97$ dB). It turns out that in the scheme, it is possible to realize the squeezed Schrodinger's cat state with probability $22 \%$ and fidelity $0.98$. In this case, it is enough to measure only one photon.

Unfortunately, the result of our exact analysis cannot be compared with the results of other works \cite{Takase2021,Ourjoumtsev_cat2007}. The fact is that our goal is to explore the possibility of the generation of squeezed Schrödinger cat states suitable for processing and transmission of quantum information. As noted earlier, for these purposes, we need states with large squeezing degree and small amplitude. In other works, devoted to generation of squeezed Schrodinger's cat states, the authors, on the contrary, studied states with large amplitudes, which are applicable in other protocols of quantum informatics. To implement such states in our scheme, we need to consider the case of measuring more than one photon.

Thus, in the presented scheme, one can realize both ordinary and squeezed Schrodinger's cat states when measuring only one photon. We have obtained that the squeezed cat states are generated with higher fidelity and higher probability. The $22 \%$ probability is quite large, because the maximum probability of a single photon measurement in this scheme is $25 \%$.

\section{Conclusion}
In this paper, we investigated the possibility of generating various non-Gaussian states using the scheme presented in Fig. \ref{fig:SVS}. We have shown that using generalized Hermite polynomials, it is possible to get explicitly the wave function of the output state. Importantly, the wave function we found depends on all scheme parameters (relative phase of the squeezed oscillators, the beam splitter transmission coefficient, and the squeezing degree) and on the number of measured photons. In other words, the wave function of the output state found by us has the highest possible generality. Such generality distinguishes our work from previous ones, in which either approximate expressions were used for calculations or specific scheme parameters were considered. The use of general expressions allowed us to comprehensively investigate the scheme in question.

For output states, we estimated the magnitude of non-Gaussianity using negativity. We have shown that all negativity values of output states can be divided into two regions: “negativity plateau” and “mountain of negativity”. In the “negativity plateau” region, the negativity does not grow with an increasing number of measured photons. Rapid growth of negativity is observed in the other area, which we call “mountain of negativity”. We have demonstrated that the type of output state negativity depends on the parity of the number of measured photons. All negativity surfaces are divided into two types. For the case when an even number of photons is measured, the value of negativity starts from zero, and for the case when an odd number of photons is measured, the negativity starts from $0.426$.

To understand the behavior of negativity, we have considered the entanglement degree of states after the beam splitter. To estimate this value, we used the van Loock-Furusawa criterion, which we adapted for our scheme. Based on this estimate, we found that the absence of an increase in the magnitude of the negativity in the “negativity plateau” region is associated with the absence of entanglement in this region. In addition, we have shown that the largest magnitude of negativity will be observed after measuring one of two oscillators which are in maximum entangled state (entangled state obtained after mixing of orthogonally squeezed oscillators on the symmetric beam splitter).

Also in the work, we have investigated the obtained non-Gaussian states. We have showed that three types of states can be got depending on the scheme parameters: Schrodinger's cat states, squeezed Schrodinger's cat states, and Fock states. As an example, we considered the possibility of generating ordinary and squeezed Schrodinger's cat states in the case of the most probable non-trivial measurement outcome, namely when a single photon is measured. We have analyzed the possibility of generating specific states, which can be used to correct quantum errors. We have showed that by using explicit expressions for the output state wave function, it is possible to find scheme parameters (squeezing degree, relative phase and the beam splitter transmission) at which the maximum of both the fidelity value and the probability of generating states will be observed. We have obtained that with the optimized parameters, the scheme can generate the Schrodinger's cat state (amplitude $\alpha=2$) with fidelity equal to $F\approx 0.88$ and probability $P\approx 18\%$. To generate the squeezed Schrodinger's cat state with $\alpha=1/2$ and $R=1$, the scheme can be optimized so that the fidelity of the output state is $F\approx0.98$ and its probability $P\approx 22\%$.

Thus, the use of explicit expressions for the output state, which depend on all the scheme parameters and on the number of measured photons, opens up great opportunities for us. With their help, it is possible to investigate the possibility of generation of certain states. The important thing is that knowing analytical expressions, one can completely optimize the scheme of generation of any given non Gaussian states.

\vspace{0.5 cm}

This research was supported by the Theoretical Physics and Mathematics Advancement Foundation "BASIS" (Grants No. 21-1-4-39-1). SBK acknowledges support by the Ministry of Science and Higher Education of the Russian Federation on the basis of the FSAEIHE SUSU (NRU) (Agreement No. 075-15- 2022-1116).



\bibliography{main}  
\appendix
\section{The Wigner function of the output state} \label{sec_WF}
To get the Wigner function of the output state, we need to evaluate the following integral:
\begin{align}
    W(x,p)=\frac{1}{\pi}\int  \Psi_{out}\left(x+y\right)\Psi_{out}^*\left(x-y\right)e^{-2iyp} dy.
\end{align}
By substituting the explicit form of the output state wave function (\ref{vfout}), we can calculate  straightforwardly:
\begin{align}
W_n(x,p)= \frac{e^{-\alpha x^2}}{\pi N(n,r,t,\varphi)}\sqrt{\left|\frac{e^r \left(e^{2 r}+i \cot \frac{\varphi }{2}\right)}{\xi}\right|^2}\int dy  H_n\left(\delta y+\delta x,\chi\right)H_n\left(-\delta^*y+\delta^*x,\chi^*\right) e^{-\alpha y^2+\beta y},
\end{align}
where for simplicity the following abbreviations will be employed:
\begin{align}
&\alpha=\Re\left[\frac{e^r-\left(1-e^{i \varphi }\right) t^2 \sinh
   r}{e^{-r}+\left(1-e^{i \varphi }\right) t^2 \sinh r}\right], \quad \beta =-2i\left(\Im \left[\frac{e^r-\left(1-e^{i \varphi }\right) t^2 \sinh
   r}{e^{-r}+\left(1-e^{i \varphi }\right) t^2 \sinh r}\right]x+p\right)\\
   &\delta=-\frac{2 e^{r} \gamma }{\xi }, \quad \chi=\tanh r \left(\frac{2(1-t^2)\sin \frac{\varphi}{2}}{\xi}-1\right),\\
   & \xi = \left(1+\left(e^{2 r}-1\right) t^2\right)\sin \frac{\varphi }{2} +i \cos \frac{\varphi }{2}, \quad \gamma  =  2 t \sqrt{1-t^2} \sin \frac{\varphi }{2} \sinh r.
\end{align}
Using a two-index Hermite polynomial, the Wigner function of the output state can be written explicitly in the following form:
     \begin{multline}
    W_n(x,p)= \frac{e^{-\alpha x^2}}{\pi N(n,r,t,\varphi)}\sqrt{\left|\frac{e^r \left(e^{2 r}+i \cot \frac{\varphi }{2}\right)}{\xi}\right|^2} \sqrt{\frac{\pi}{\alpha}}\exp \left[\frac{\beta^2}{4\alpha}\right]\\
    H_{n,n}\left(\delta x+\frac{\delta \beta}{2\alpha},\chi+\frac{\delta^2}{4\alpha},\delta^*x+\frac{\delta^* \beta ^*}{2\alpha},\chi^*+\frac{\delta^{*2}}{4\alpha}\Big|-\frac{\left|\delta\right|^2}{2\alpha}\right).
 \end{multline}
For simplicity, we rewrite the two-index Hermite polynomial in  term of a generalized Hermite polynomial (\ref{tiHp}). The Wigner function can be further simplified. It is observed that
 \begin{multline}
 W_n\left(x,p\right)=\frac{1}{N\left(n,r,t,\varphi\right)}\sqrt{\frac{\cot^2 \frac{\varphi}{2}+e^{4r}}{\pi\left(\gamma^2+1\right)}}  \exp \left[-\frac {p^2\frac{|\xi|^2}{e^{2r}}+x^2\frac{ \left(\gamma ^2+1\right)^2+\eta^ 2}{e^{-2r}|\xi|^2}+2xp\eta}{\gamma
   ^2+1}\right]\\
   \sum \limits _{k=0}^{n} \binom{n}{k}^2k!\left(-\frac{2 \gamma ^2}{\gamma ^2+1}\right)^k\left|H_{n-k}\left(-\frac{2 \gamma}{\gamma ^2+1}\left(\frac{\xi -2 t^2 \sin \frac{\varphi}{2}\sinh 2 r}{e^{-r}}x+i\frac{\xi}{e^r}p\right),-\frac{t^2+e^{i \varphi } \left(1-t^2\right)}{\left(\gamma ^2+1\right)\coth r}\right)\right|^2,
\end{multline}
where we have defined $\eta= t^2\sin \varphi \sinh 2 r$.

\section{The separability criterion} \label{sec_EC}
 In the paper we use the van Loock–Furusawa criterion, adapted to our scheme, to determine the entanglement degree. In a general case, this criterion imposes a condition on the variances of nullifier operators characterized by the system. According to this criterion, a system is entangled if the quantum fluctuations of the operators satisfy the inequality. 

The two nullifier operators completely defined our system can be written as
\begin{align} \label{null_1}
    &\hat{N}_1=\hat{u}_1+\hat{u}_2,\\
     &\hat{N}_2=\hat{v}_1+\hat{v}_2, \\
     &\hat{u}_1=t\hat{X}_1, \quad \hat{u}_2=\sqrt{1-t^2}\hat{X}_2,\\
     &\hat{v}_1=-\sqrt{1-t^2}\left(\hat{X}_1\cos \frac{\varphi}{2}+\hat{Y}_1\sin \frac{\varphi}{2}\right),\\
     &\hat{v}_2=t\left(\hat{X}_2\cos \frac{\varphi}{2}+\hat{Y}_2\sin \frac{\varphi}{2}\right) \label{null_2}
\end{align}
where $\hat{X}_i$ and $\hat{Y}_i$ are the $i$-th- the quadratures of the entangled state at the output of the beam splitter, $\varphi/2$ is the rotation angle on the phase plane of the second oscillator versus the first one. The operators $\hat{u}_i$ and $\hat{v}_i$ are a linear combination of the quadratures of the $i$-th subsystem. The two-particle state is separable, then the following condition on the sum the variances of nullifier operators \cite{Furusawa_2003} is satisfied
\begin{align}
    \langle \delta \hat{N}_1^2\rangle+\langle \delta \hat{N}_2^2\rangle \geqslant \left|\left[\hat{u}_1,\hat{v}_1\right]\right|+\left|\left[\hat{u}_2,\hat{v}_2\right]\right|.
\end{align}
By substituting the operators (\ref{null_1})-(\ref{null_2}) into this condition, as well as the quadratures of the output of the beam splitter, we obtain the following separability criterion of the output state:

\begin{align}
     \frac{1}{2}e^{-2r}+\frac{1}{2}e^{-2r} \geqslant 2\sqrt{1-t^2}t \sin \frac{\varphi}{2}.
\end{align}
If this condition is not satisfied then the output state is not separable, that is, it is entangled. Thus, the ultimate condition of the entanglement in our system is given by the following inequality:

\begin{align}
    \frac{2\sqrt{1-t^2}t \sin \frac{\varphi}{2}}{e^{-2r}}>1.
\end{align}
In this case, the higher the value in the LHS of the inequality, the more entangled the state is. 
\end{document}